\documentclass{emulateapj}
\usepackage{rotating}
\usepackage{setspace}

\slugcomment{Accepted for publication in \aj}

\shorttitle{{\it Kepler}-INT Survey (KIS)}
\shortauthors{Greiss et al.}

\begin{document}


\title{Initial data release of the {\it Kepler}-INT Survey$^{\dagger}$}
%


\author{S. Greiss\altaffilmark{1}, D. Steeghs\altaffilmark{1}, B. T. G\"ansicke\altaffilmark{1}, E. L. Mart\'in\altaffilmark{2}, P. J. Groot\altaffilmark{3}, M. J. Irwin\altaffilmark{4}, E. Gonz\'alez-Solares\altaffilmark{4}, R. Greimel\altaffilmark{5}, C. Knigge\altaffilmark{6}, R. H. \O stensen\altaffilmark{7}, K. Verbeek\altaffilmark{3}, J. E. Drew\altaffilmark{8}, J. Drake\altaffilmark{9}, P. G. Jonker\altaffilmark{3, 9, 10}, V. Ripepi\altaffilmark{11}, S. Scaringi\altaffilmark{3}, J. Southworth\altaffilmark{12}, M. Still\altaffilmark{13,14}, N. J. Wright\altaffilmark{9}, H. Farnhill\altaffilmark{8}, L. van Haaften\altaffilmark{3}, S. Shah\altaffilmark{3}}
\altaffiltext{1}{Department of Physics, Astronomy and Astrophysics group, University of Warwick, CV4 7AL, Coventry, U.K.}
\altaffiltext{2}{INTA-CSIC Centro de Astrobiolog\'ia, Carretera de Ajalvir km 4, 28550 Torrej\'on de Ardoz, Spain}
\altaffiltext{3}{Department of Astrophysics/IMAPP, Radboud University Nijmegen, P.O.Box 9010, 6500 GL, Nijmegen, The Netherlands}
\altaffiltext{4}{Cambridge Astronomy Survey Unit, Institute of Astronomy, University of Cambridge, Madingley Road, CB3 0HA, Cambridge, U.K.}
\altaffiltext{5}{Institut f\"ur Physik, Karl-Franzen Universit\"at Graz, Universit\"atsplatz 5, 8010 Graz, Austria}
\altaffiltext{6}{School of Physics and Astronomy, University of Southampton, Southampton, Hampshire, SO17 1BJ, U.K.}
\altaffiltext{7}{Instituut voor Sterrenkunde, KU Leuven, Celestijnenlaan 200D, 3001 Leuven, Belgium}
\altaffiltext{8}{Centre for Astrophysics Research, University of Hertfordshire, College Lane, Hatfield, AL10 9AB, U.K.}
\altaffiltext{9}{Harvard-Smithsonian Center for Astrophysics, 60 Garden Street, Cambridge, MA 02138, USA}
\altaffiltext{10}{SRON, Netherlands Institute for Space Research, Sorbonnelaan 2, 3584 CA, Utrecht, The Netherlands}
\altaffiltext{11}{INAF-Osservatorio Astronomico di Capodimonte, via Moiariello 16, Naples, I-80131, Italy}
\altaffiltext{12}{Astrophysics Group, Keele University, Newcastle-under-Lyme, ST5 5BG, U.K.}
\altaffiltext{13}{NASA Ames Research Center, M/S 244-40, Moffett Field, CA 94035, USA}
\altaffiltext{14}{Bay Area Environmental Research Institute, Inc., 560 Third St. West, Sonoma, CA 95476, USA}
\email{s.greiss@warwick.ac.uk}

%
%
%
%



\begin{abstract}
This paper describes the first data release of the {\it Kepler}-INT Survey (KIS), that covers a 116 deg$^{2}$ region of the Cygnus and Lyra constellations. The {\it Kepler} field is the target of the most intensive search for transiting planets to date. Despite the fact that the {\it Kepler} mission provides superior time series photometry, with an enormous impact on all areas of stellar variability, its field lacks optical photometry complete to the confusion limit of the {\it Kepler} instrument necessary for selecting various classes of targets. For this reason, we follow the observing strategy and data reduction method used in the IPHAS and UVEX galactic plane surveys in order to produce a deep optical survey of the {\it Kepler} field.  This initial release concerns data taken between May and August  2011, using the Isaac Newton Telescope on the island of La Palma. Four broadband filters were used, $U, g, r, i$, as well as one narrowband one, H$\alpha$, reaching down to a 10$\sigma$ limit of $\sim$ 20$^{th}$ mag in the Vega system. Observations covering $\sim$ 50 deg$^{2}$, thus about half of the field, passed our quality control thresholds and constitute this first data release. We derive a global photometric calibration by placing the KIS magnitudes as close as possible to the {\it Kepler} Input Catalog (KIC) photometry. The initial data release catalogue containing around 6 million sources from all the good photometric fields is available for download from the KIS webpage$^{\dagger}$, as well as via MAST$^{\star}$.
\end{abstract}


\section{Introduction}

We present an initial data release of the {\it Kepler}-INT Survey (KIS)\let\thefootnote\relax\footnotetext{$\dagger$ KIS webpage url: $\mathrm{www.astro.warwick.ac.uk/research/kis/}$.}\let\thefootnote\relax\footnotetext{$\star$ KIS magnitudes can be retrieved using the MAST Enhanced Target Search Page [$\mathrm{http://archive.stsci.edu/kepler/kepler_fov/search.php}$], as well as via Casjobs at MAST [$\mathrm{http://mastweb.stsci.edu/kplrcasjobs/}$]}. This paper describes optical observations carried out on the Isaac Newton Telescope (INT), covering about half of the {\it Kepler} field down to $\sim$ 20$^{th}$ magnitude. A short description of the {\it Kepler} mission and the INT is given in this Section of the paper. Section 2 describes the INT observations and data products, while Section 3 explains the photometric calibration of the data using the {\it Kepler} Input Catalog (KIC, \citealt{brownetal11}). In Section 4, we provide a description of the catalogue. 

\subsection{{\it Kepler} mission}

The {\it Kepler} mission's \citep{boruckietal10} main goal is to discover Earth-size planets within the habitable zones of Sun-like stars. NASA's {\it Kepler} spacecraft, which was launched in March 2009, contains a differential broadband optical (4,200 - 9,000$\mathrm{\mathring{A}}$) CCD array with a wide field of view (FoV) of 116 deg$^{2}$, mounted on a modified 0.95m Schmidt telescope continuously observing a region in the Cygnus and Lyra constellations. Due to the onboard storage and telemetry bandwidth limitation, only 170,000 sources, out of the millions present within the FoV, can be observed and downloaded to Earth at any given time. Therefore, the targets must be selected prior to the observations. \\

 {\it Kepler} provides uninterrupted time series photometry that is superior to any previous ground-based study. Although {\it Kepler} was designed for the detection of exoplanets, its high-quality light curves hold an enormous potential for other astrophysical domains such as asteroseismology \citep{chaplinetal10}, stellar activity \citep{basrietal11}, star spot monitoring \citep{llamaetal12}, eclipsing and close binary systems \citep{prsaetal11, coughlinetal11, bloemenetal11}, gyrochronology \citep{meibometal11}, accreting white dwarfs \citep{fontaineetal11, stilletal10, woodetal11}, the study of RR Lyrae stars \citep{benkoetal10, nemecetal11} as well as systems showing stochastic behaviour in the variability of their fluxes \citep{mushotzkyetal11, scaringietal12}. {\it Kepler} data has also enabled the first determination of radial velocity amplitudes of binary systems through Doppler boosting \citep{vankerkwijketal10}.
 
{\it Kepler} operates two types of observation modes: the short (one minute) and long (30 minutes) cadence modes. The Guest Observer (GO) program offers a yearly opportunity for the observation of 5,000 long cadence targets per quarter and 40 short cadence targets per month, through a peer-reviewed competition, which is open for all astrophysical domains. Every 3 months, the {\it Kepler} mission also offers the opportunity for a few dozen targets to be observed through Director's Discretionary Time (DDT) Proposals. Finally, every quarter, the {\it Kepler} Asteroseismology Science Consortium (KASC) can bid for more than $\sim$ 1,700 targets in order to study stellar pulsations. \\

It is therefore clear that the short cadence mode slots are very limited, thus a target must be well studied from the ground in order to justify required time with {\it Kepler}. In order to observe candidate planet hosts, mainly G-M type main-sequence stars, the {\it Kepler} team created the Stellar Classification Project (SCP), with a main goal to prevent the selection of non-main-sequence stars, by providing important stellar parameters (radius, effective temperature, apparent magnitude, etc) of the sources in the {\it Kepler} FoV. A photometric study of the {\it Kepler} field, mainly using $griz$ broadband filters was produced and stored in what is known as the {\it Kepler} Input Catalog \citep{brownetal11}. Since the main purpose of the KIC was to pre-select bright solar-like stars in order to detect Earth-like planets around them, the reliable depth of this survey is $g$ $\sim$ 16 mag and there was no need to include a filter bluer than the $g$-band. However, it is clear that many fainter objects within the {\it Kepler} FoV, which cannot be selected using KIC data, are of interest to non-exoplanet science such as cataclysmic variables \citep{woodetal11}, pulsating white dwarfs \citep{ostensen11-2, hermes11} and active galactic nuclei \citep{mushotzkyetal11}.
 
 The GO and KASC programs to date show that there is a large interest in fainter and bluer objects. In order to pre-select other, rarer types of targets such as hot, young, or active stars, white dwarfs or subdwarfs, and accreting objects, a deeper optical survey of the {\it Kepler} field, including a  filter bluer than $g$, is required. Also, the addition of an H$\alpha$ filter would be useful to detect emission line objects, as well as strong H$\alpha$ deficit sources such as hydrogen-rich white dwarfs. Therefore, the INT Photometric H$\alpha$ Survey of the Northern Galactic Plane (IPHAS, \citealt{drewetal05-1}) and the UV-Excess Survey  of the Northern Galactic Plane(UVEX, \citealt{grootetal09}) collaborations made use of their available data reduction pipeline and observation strategy to obtain a homogeneous $Ugri$ and H$\alpha$ catalogue of the {\it Kepler} FoV, down to $\sim$ 20$^{th}$ mag in all five filters. All magnitudes are given in the Vega system \citep{morgan+johnson53}. We have named this effort the Kepler-INT Survey (KIS). KIS should be useful not only because it can identify UV-excess objects and H$\alpha$ emitters, but also because it goes much deeper than KIC. Even though other collaborations are also conducting optical surveys of the {\it Kepler} field, such as the UBV Photometric Survey of the {\it Kepler} field \citep{everettetal12}, only KIS provides the critical deep $U$-band and H$\alpha$ imaging.\\

\subsection{Survey imaging with the Isaac Newton Telescope}

The 2.5m Isaac Newton Telescope (INT) is located in the Roque de los Muchachos Observatory on La Palma. The Wide Field Camera (WFC), mounted in its prime focus, is an optical imager consisting of 4 anti-reflective-coated 2048 $\times$ 4096 pixel CCDs, arranged in an L-shape. It has a pixel scale of 0.333 arcsec and a field of view of 0.29 deg$^{2}$ \citep{gonzalez-solares08}. \\

Four broadband filters ($Ugri$) and one narrowband filter (H$\alpha$) were used to obtain the INT data. The filter characteristics are provided in Table \ref{int-filters}. Unlike $g, r$ and $i$ which are SDSS-like filters, the $U$-band is a non-standard $U$ filter and it is affected by the CCD detector response dropping towards its blue edge. For more information on the $U$ filter,  see \cite{verbeeketal12}. \\
 
\begin{table}
\caption{Filter parameters of INT observations \citep{gonzalez-solares08, grootetal09} \label{int-filters}}
\centering
\begin{tabular}{c c c}
\hline \hline
\noalign{\smallskip}
Filters & Central wavelength ($\mathrm{\mathring{A}}$) & FWHM ($\mathrm{\mathring{A}}$)\\
\hline
$U$ & 3581 & 638 \\
$g$ & 4846 & 1285 \\
$r$ & 6240 &1347 \\
$i$ & 7743 & 1519 \\
H$\alpha$ & 6568 & 95 \\
\hline
\end{tabular}
\end{table}

The wealth of the available IPHAS and UVEX data has been used to develop selection methods to detect objects of special interest such as H$\alpha$ emitters \citep{withametal06-1}, cataclysmic variables \citep{withametal07}, planetary nebulae \citep{viironenetal09}, symbiotic stars \citep{corradietal08}, early-A stars \citep{drewetal08}, extremely red stellar objects, including mainly Asymptotic Giant Branch stars, and S-type stars \citep{wrightetal08, wrightetal09}, very low-mass accreting stars and brown dwarfs \citep{valdivielsoetal09} and UV-Excess sources \citep{verbeeketal12}. Candidates were primarily selected through the use of colour-colour diagrams. Their nature and the efficiency of the associated selection methods were then confirmed using spectroscopic data. Surveys such as IPHAS and UVEX have enabled the development of automated searches of a large number of unusual and `exotic' objects. 

\section{INT observations and data}

\subsection{Observations}

As our data processing recipe is identical to that of the IPHAS and UVEX surveys, we refer to  \cite{drewetal05-1}, \cite{gonzalez-solares08} and \cite{grootetal09} for details. The observing strategy consists of dividing the entire survey region into fields, each of them corresponding to the area of the WFC's FoV. A five percent overlap is included between adjacent pointings. Also, in order to cover the gaps between the four detectors, comprising $\sim$ 12 arcmin$^{2}$, each field is observed in pairs with an offset of 5 arcmin North and 5 arcmin East between the pointings. This leads to at least two detections of most objects observed.

In order to balance the survey progress with the calibration quality of the data, approximately five observations of standards fields are taken throughout each night. These observations allow us to derive accurate zero-point magnitudes (ZPs) for each broadband filter per night without using too much of the allocated time on the telescope. The zero-point RMS of each night allows one to assess whether a night is considered `photometric' \citep{gonzalez-solares08}.\\

\subsection{Data}

The data processing is described in detail in Section 3 of \cite{gonzalez-solares08}. The final data products consist of band-merged catalogues with equatorial positions tied to 2MASS (Two Micron All Sky Survey, \citealt{skrutskieetal06-1}), Vega magnitudes and errors in all five filters and morphological flags (see Table \ref{morph}). Further information on each detected object, such as CCD pixel coordinates in each waveband and the CCD in which the source was detected, are also provided in the catalogues. The astrometric precision of the end product is better than 100 mas across all four CCDs \citep{gonzalez-solares08}.

In the UVEX data reduction pipeline, the $U$-band zero-point (ZP) magnitudes are tied to the $g$-band ones with a fixed offset of (ZP$_{g}$ - ZP$_{U}$) = 2.1 mag, similarly to the case of the H$\alpha$ ZPs in IPHAS \citep{drewetal05-1} which are tied to the $r$-band via a fixed offset of 3.14 mag. The nightly $g$-band zero-points are derived from the standards observed throughout each night. However, in the KIS, we depart from the UVEX strategy in the $U$-band by using actual standard star ZPs to obtain $U$-band magnitudes, in the same way as we do for $g, r$ and $i$ (see Section \ref{calibration} for more details).  The H$\alpha$ zero-points for each night remain tied to the $r$-band ones, as is done in IPHAS \citep{drewetal05-1}, since there are no H$\alpha$ standards available.
 
 A bubble in the $U$ filter was discovered that was visibly affecting a corner of the $U$-band images, and a red leak is also known to exist in the filter \citep{verbeeketal12}. The bubble was fixed on the 15$^{th}$ of June 2011. However, Calima - dust winds which originate in the Saharan desert - was strongly present during that period of the observations. This affected the pre-June 20$^{th}$ $U$-band data in particular. The derived $U$-band ZPs taken from the standards observations of those nights have an RMS of $\sim$ 0.3 mag, about three times larger than the quality control threshold set for this survey and are thus not included in this release. \\
 
\begin{table}
\caption{Morphological flags \label{morph}}
\centering
\begin{tabular}{c c}
\hline \hline
Flags & Definition\\
\hline
-9 & saturated \\
-8 & poor match \\
-7 & contains bad pixels \\
-1 & stellar \\
-2 & probably stellar \\
-3 & compact but probably not stellar \\
1 & non-stellar (e.g. a galaxy) \\
0 & no detection\\
\hline
\end{tabular}
\end{table}

\subsection{Quality control flags}
\label{quality-control}

\begin{figure}
\includegraphics[width=9cm, height=9cm]{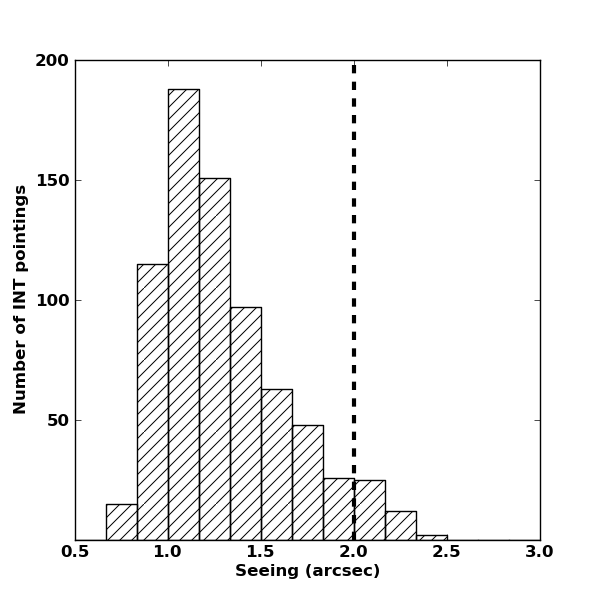}
\caption{$r$-band seeing in arcseconds of all INT pointings. Fields observed under seeing conditions worse than 2 arcsec were not included in the initial data release catalogue. \label{seeing-r}}
\end{figure}

 During the 2011 observing season, a total of 742 INT pointings, consisting of fields and offsets, were observed. However, not all of them pass the quality control threshold set for this survey. We only select fields which were observed under reasonably clear conditions where the RMS on the derived nightly zero-points must be smaller than 0.10 mag. Additional quality control tests related to the observing conditions include selecting pointings which have $r$-band seeing~$<$ 2~arcsec and $r$ and $g$-band sky background values~$<$~2000 ADUs, to remove observations done too close to the moon. The distribution of seeing in the $r$-band, for all 742 pointings, is shown in Figure \ref{seeing-r}. Also, we use an additional measurement, the ellipticity, which is a detector-averaged PSF property that flags any tracking and focussing issues of the telescope that were possibly encountered on a given night. It must not be confused with a shape measurement for each source. We keep fields with mean $r$-band ellipticity values below 0.2 and any larger value would trigger a re-observation.   \\

\begin{figure}
\includegraphics[trim= 1.6cm 0.5cm 0cm 2.3cm, clip, width=9.5cm]{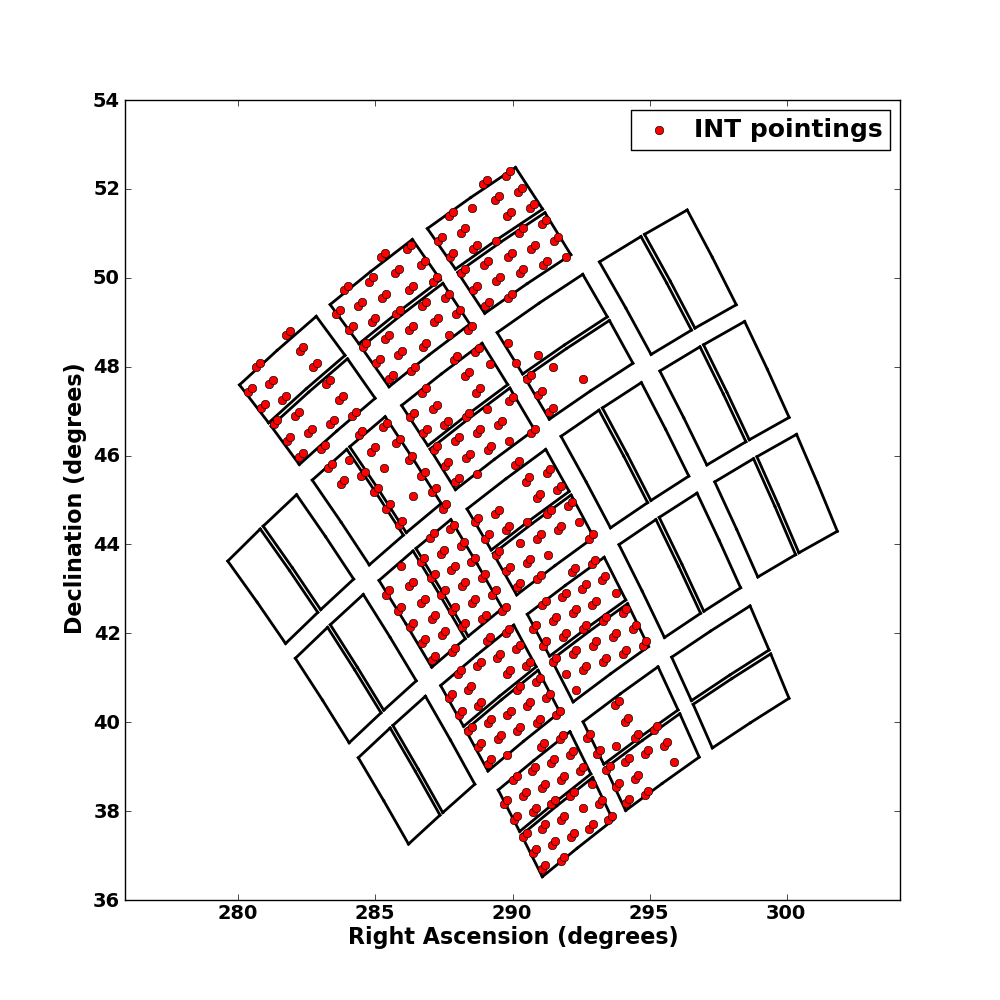}
\caption{INT coverage of {\it Kepler} fields. The red circles correspond to the centres of the INT pointings which passed our quality control tests and are part of this release. The boundaries of the sky footprints of the CCDs on the {\it Kepler} satellite are also shown in black. \label{int-cov}}
\end{figure}

 Out of all 742 observed fields, 511 pointings pass our quality control tests, which is equivalent to $\sim$ 70\% of the total number of pointings observed so far and $\sim$ 50\% of the {\it Kepler} field. This number includes an additional quality control criterion described in the following Section. In Figure \ref{int-cov}, we plot the centres of these `good' INT pointings on the {\it Kepler} FoV. We indicate the boundaries of the sky footprints of the CCDs on the {\it Kepler} satellite. Note that our images include the gaps in between the {\it Kepler} CCDs. Thus not all objects in our catalogues will land on-chip.\\

\section{Photometric calibration}
\label{calibration}

 For the KIS catalogue, we start by calibrating $Ugri$ to the standards observed each night by taking the average nightly ZPs in each filter, while the H$\alpha$ ZP is tied to the nightly $r$-band ZP by a fixed offset. This calibration can introduce ZP errors if the night is not reasonably clear since all pointings in a given night employ the same ZPs. As mentioned previously, we reject the nights that have ZP RMS deviations larger than 0.10 mag.  \\

Given the existence of the well-calibrated KIC catalogue, which has served as the principal survey for selecting {\it Kepler} targets, we decided to tie our absolute photometric calibration to the KIC broad-band magnitudes. 

KIC contains over 13 million detected objects. A full explanation of the catalogue production can be found in \cite{brownetal11}, but we provide a brief description of it here. The KIC photometric data were placed as close as possible to the Sloan photometric system \citep{brownetal11}, by selecting 8 fields outside the {\it Kepler} FoV, which overlap with SDSS DR1 \citep{stoughtonetal02-1} and which are used as photometric standards. Spanning a wide range of RA around the {\it Kepler} field, 316 primary standard stars were chosen. Each night, standards were taken on an hourly basis in order to calculate the transformations between the KIC and SDSS magnitudes. A specially designed pipeline was used to reduce the image data to catalogues of star positions and apparent magnitudes. The photometric calibrations were done using the time-averaged extinction-corrected magnitudes from the standards stars \citep{brownetal11}. The photometric precision of the KIC sources is expected to be $\sim$ 1.5\%. However, it is important to note that out of the $\sim$ 13 million detected objects, less than 3 million have $g$-band magnitudes~$<$~16. The rest are either fainter than 16$^{th}$ mag or are not provided with a magnitude value but only with their coordinates. \\

Given that the astrometry of both the KIS and KIC catalogues are based on well-resolved CCD data, we used a matching radius of 1 arcsec. Additional information on the KIC astrometry can be found on the {\it Kepler} webpage$^\star$\let\thefootnote\relax\footnotetext{$^\star$http://keplergo.arc.nasa.gov/Documentation.shtml}. Since the KIC photometry is based on the AB system \citep{oke+gunn83} and the INT data is all in the Vega magnitude system, we converted the KIC data to the Vega system using the transformations from \cite{gonzalez-solares11}:
\begin{eqnarray}
\label{colour-tran}
 U_\textrm{\tiny WFC} = u_\textrm{\tiny SDSS} - 0.833 - 0.009 \times (u_\textrm{\tiny SDSS} - g_\textrm{\tiny SDSS})  \nonumber \\
 g_\textrm{\tiny WFC} = g_\textrm{\tiny SDSS} + 0.060 - 0.136 \times (g_\textrm{\tiny SDSS} - r_\textrm{\tiny SDSS})  \nonumber \\
 r_\textrm{\tiny WFC} = r_\textrm{\tiny SDSS} - 0.144 + 0.006 \times (g_\textrm{\tiny SDSS} - r_\textrm{\tiny SDSS})  \nonumber \\
 i_\textrm{\tiny WFC} = i_\textrm{\tiny SDSS} - 0.411 - 0.073 \times (r_\textrm{\tiny SDSS} - i_\textrm{\tiny SDSS})  
 \end{eqnarray}
 
 \begin{figure}
\includegraphics[trim= 1.6cm 0cm 0cm 1cm, clip, width=9.5cm]{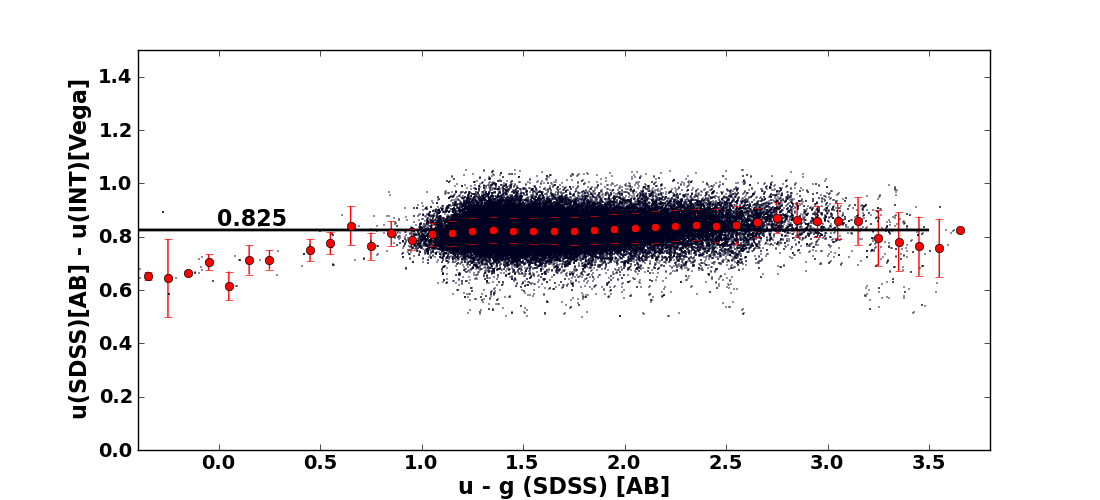}
\caption{Difference between $u$ from SDSS (AB system) and $U$ from KIS (Vega system) against ($u - g$) from SDSS (AB system), showing that the transformations from AB to Vega (and vice-versa) are more colour-dependent when looking at non-main-sequence stars. The red points correspond to the medians of ($u - U$) over 0.1 magnitude bins in ($u - g$)[AB]. The median of ($u - U$) is 0.825, a value close to the fixed term found in Equations \ref{colour-tran}. The data points with no error bars simply mean that only one data point was used to determine the median. \label{u_vs_ug}}
\end{figure}

We stress that these transformations are reliable for main-sequence stars but are not to be trusted for blue objects which have a negative ($U - g$) colour. In order to verify this, we cross-match the KIS data with Sloan Digital Sky Survey (SDSS, \citealt{abazajianetal09}) and select matches with $r$-band magnitudes ranging from 15 to 18 mag. Only $\sim$ 25\% of the KIS pointings overlap with SDSS and therefore this test is only used to determine the range over which these transformations are valid. The transformations taken from \cite{gonzalez-solares11} were derived using a more robust algorithm. 
 
 In our test, we plot the difference between SDSS and KIS magnitudes in $U$ against ($u - g$) colours from SDSS (see Figure \ref{u_vs_ug}). We bin the data in colour bins of 0.1 mag and calculate the median of the difference between the SDSS and KIS magnitudes. These values correspond to the red circles. The error bars are the standard deviations of the binned data. As we can see, the transformations provided by \cite{gonzalez-solares11} are confirmed for objects within 1~$<$~($u-g$) [AB]~$<$~3. At both the red and blue ends of the plots, the data points do not follow the linear fit for the main locus of stars. For the purpose of our photometric calibration, this is not an issue.\\

\begin{figure}
\includegraphics[trim= 0.5cm 0cm 0cm 1cm, clip, width=9cm]{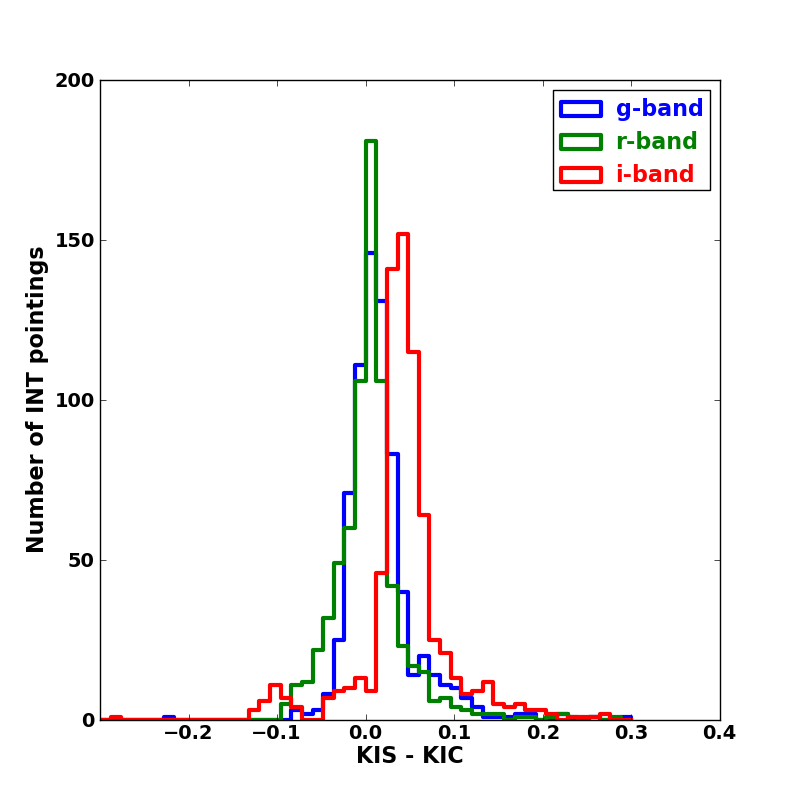}
\caption{Distribution of $\Delta g$, $\Delta r$ and $\Delta i$, for all pointings, where $\Delta \{filter\}$ is the offset between the KIC and KIS magnitudes (see Section \ref{calibration}). As one can see, the offsets rarely exceed 5\%, and fields with offsets $>$ 0.2 mag from the median are rejected. \label{dist-offsets}}
\end{figure}

As mentioned earlier, we use the KIC to calibrate the photometric data of KIS. By placing the KIS ZPs as close to the KIC ones as possible, we can improve our photometric calibration on a pointing by pointing basis. We calculate the difference between the KIC magnitudes and the KIS ones for sources with $g$-band magnitudes between 13 and 15$^{th}$ mag. These limits were chosen because KIS magnitudes smaller than $\sim$ 12$^{th}$ mag become less reliable due to saturation and the photometric accuracy of KIC deteriorates above $\sim$ 16$^{th}$ mag. We plot the distribution of the offsets in the $g, r$ and $i$ bands in Figure \ref{dist-offsets} ($\Delta g$, $\Delta r$ and $\Delta i$). The median values of these offsets corresponding to the centre of their distributions, as well as the standard deviations of these distributions, are given in Table \ref{med-off}. As can be seen, the values of these offsets were typically a few percent, rarely exceeding 5\% for fields passing the quality control threshold. The applied offset values are provided in the final catalogue of the {\it Kepler}-INT Survey.

We also make use of the KIC to set an additional quality control criterion: we only select fields which have a median offset with respect to KIC in each waveband within $\pm$ 0.2 mag of the values given in Table \ref{med-off}.  Only 34 fields fail this criterion. In general, the systematic offset between the KIC and KIS magnitudes are within the range of 0.1 mag. \\

\begin{table}
\caption{\label{med-off} $\Delta \{filter\}$ is the median offset between the KIC and KIS magnitudes (see Section \ref{calibration}). The standard deviations of the distributions in each filter are also given here ($\sigma_{\{filter\}}$). The distributions are shown in Figure \ref{dist-offsets}.}
\centering
\begin{tabular}{c c c c}
\hline \hline
 & $g$ & $r$ & $i$ \\
$\Delta \{filter\}$ & 0.012 & 0.004 & 0.044 \\
$\sigma_{\{filter\}}$ & 0.040 & 0.042 & 0.053 \\
\hline
\end{tabular}
\end{table}

 The KIC contains $g$, $r$ and $i$ - band magnitudes for a large number of sources. For these bands, our photometric corrections are thus simply:
\begin{eqnarray}
\Delta g = g_\textrm{\tiny WFC} - g_\textrm{\tiny KIC}  \nonumber \\
\Delta r = r_\textrm{\tiny WFC} - r_\textrm{\tiny KIC}  \nonumber \\
\Delta i = i_\textrm{\tiny WFC} - i_\textrm{\tiny KIC}  
\end{eqnarray}

 In order to have a more accurate calibration, we calculated these offsets for each WFC CCD separately. In general, all four CCDs behaved the same way. For each KIS field, we calculate the median of the offsets of all matched sources in each passband for each WFC CCD (for instance, median($\Delta g_\textrm{\tiny CCD1}$), median($\Delta g_\textrm{\tiny CCD2}$),median($\Delta g_\textrm{\tiny CCD3}$), median($\Delta g_\textrm{\tiny CCD4}$), and similarly for the $r$ and $i$ bands). \\

 As KIC lacks $u$-band data, the KIS $U$-band can not have an absolute calibration like in the case of $g, r$ and $i$. For two main reasons, we decided to use the $g$-band offset also for $U$: to conserve the $U - g$ colours of the KIS sources and to minimise any wavelength dependent effects by using the nearest KIC broadband, which is the $g$-band. This introduces a systematic error in the $U$-band photometry of up to $\sim$ 0.05 mag.
 
 In summary, to calibrate the INT photometry, we applied the following equations to the sources in each INT pointing and CCD:
\begin{eqnarray}
U'_\textrm{\tiny{WFC}} = U_\textrm{\tiny WFC} - \Delta g_\textrm{\tiny CCD\#}  \nonumber \\
g'_\textrm{\tiny WFC} = g_\textrm{\tiny WFC} - \Delta g_\textrm{\tiny CCD\#}  \nonumber \\
r'_\textrm{\tiny WFC} = r_\textrm{\tiny WFC} - \Delta r_\textrm{\tiny CCD\#}  \nonumber \\
i'_\textrm{\tiny WFC} = i_\textrm{\tiny WFC} - \Delta i_\textrm{\tiny CCD\#}  \nonumber \\
H\alpha_\textrm{\tiny WFC}' = H\alpha_\textrm{\tiny WFC} - \Delta r_\textrm{\tiny CCD\#}  
\end{eqnarray}
where the prime indicates the calibrated magnitudes. \\

\section{Catalogue description}

\begin{figure}
\includegraphics[trim= 0.75cm 0.5cm 0cm 1.5cm, clip, width=9.5cm]{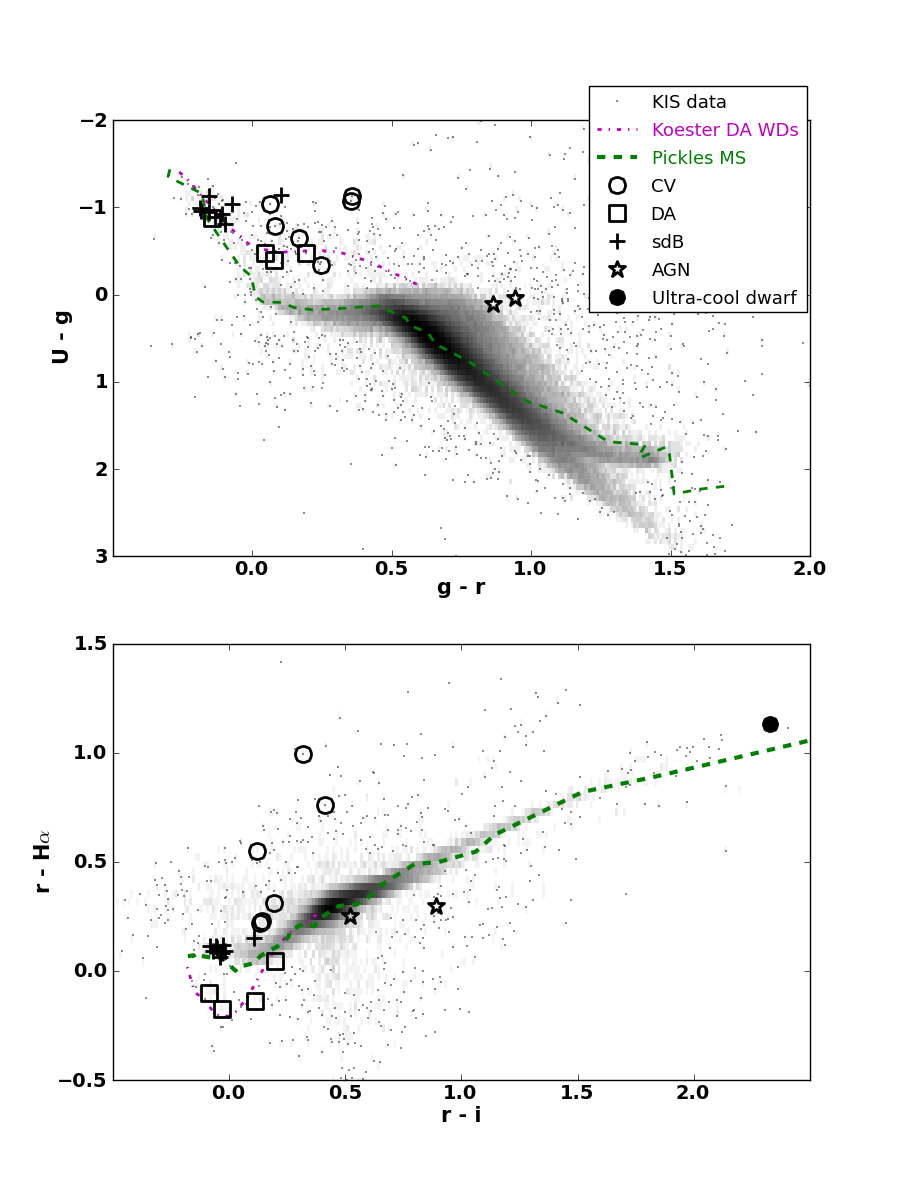}
\caption{Colour-colour diagrams of some of the published pulsators, white dwarfs, CVs, ultra-cool dwarfs and AGNs in the {\it Kepler} field. The Pickles tracks are taken from Drew et al. (2005) in the lower panel, and from Groot et al. (2009) in the top panel. The magenta tracks in the both panels correspond to Koester models of DA WDs with constant surface gravity, log $g$ = 8, taken from Verbeek et al. (2012). The grey scale and black points are stellar objects taken from the KIS catalogue which have photometric errors smaller than 0.02 mag in all five filters, as well as $r$-band magnitudes between 12 and 20$^{th}$ mag. The grey scale shows the densest region of the colour-colour diagrams using a logarithmic scale. \label{ccd}}
\end{figure}

 The 511 fields that were classified as reliable and observed during reasonably clear nights are all observed starting from the 21$^{st}$ of June 2011. We plotted individual ($U - g$, $g - r$) and ($r - \mathrm{H}\alpha$, $r - i$) colour-colour diagrams of each INT pointing in order to assess the state of the data by considering Pickles \citep{pickles98} main-sequence tracks, as well as DA (hydrogen-rich) white dwarf tracks from Koester models  \citep{koester10}, taken from \cite{drewetal05-1, grootetal09} and \cite{verbeeketal12}. A stacked colour-colour diagram of all the INT pointings is shown in Figure \ref{ccd}. All the synthetic colours are available in the Vega system from \cite{drewetal05-1} for  ($r - \mathrm{H}\alpha$, $r - i$) colours of main-sequence stars, and \cite{grootetal09} for their ($U - g$, $g - r$) colours. The DA white dwarfs colours were taken from \cite{verbeeketal12}. As can be seen in Figure \ref{ccd}, the model tracks and the stellar content of KIS are in good agreement. \\

With all reliable pointings at hand, we produced the KIS initial release catalogue containing $\sim$ 6 million sources. Out of these sources, $\sim$ 2.1 million of them are also unique detections in KIC. However, only $\sim$ 280,000 of those $\sim$ 2.1 million objects have KIC magnitudes between 12 and 16$^{th}$ mag. Therefore, we provide more `reliable' magnitudes to $\sim$ 1.3 million sources already existing in KIC. With KIS, we detect an additional $\sim$ 2.4 million unique objects in the field, since around 25\% of the sources in KIS have another detection from overlapping regions between different pointings. \\

 We produce two versions of the catalogue: a standard version and an extended one. A description of the columns of the catalogue is given in Table \ref{col-desc} and an example of a few lines taken from the standard version is shown in Table \ref{cat}. It contains the positions of the sources in degrees, their magnitudes and errors, as well as their morphological classes in each waveband (see Table \ref{morph} for more details). We also give each KIS source an ID, found in the first column of the tables. It simply corresponds to the `KISJ' prefix, followed by the object's KIS coordinate given in sexagesimal notation.  Finally, when a KIS object had a match in the {\it Kepler} Input Catalog within 1 arcsec, we added the {\it Kepler} ID of that match in the final column of the table. If no match was found, the {\it Kepler} ID is equal to 0.

 The extended catalogue contains further information on each source, such as the CCD in which it was detected, its CCD pixel coordinates, the seeing, ellipticity and modified Julian date in each filter and the offsets between the KIS magnitudes and KIC magnitudes which were used to calibrate the KIS catalogue. The seeing and ellipticity values given for each source are average values for the given INT pointing in which the object was detected. 

 Most sources had two detections, therefore the magnitude and errors provided in the catalogue are mean values of the magnitudes and errors, calculated from the magnitudes and errors of detections of the same source found in an INT and its paired field. The reason why we allow for a mean value to be calculated in this case is because the paired fields are observed one after the other, under very similar conditions. We also compute the RMS deviation of the magnitudes for each source in order to compare the difference between the magnitudes in both detections, within their error bars. If the value of the calculated RMS deviation is large compared to the errors, it would imply either short timescale variability or non-ideal observing conditions. In the case of a single detection, the RMS deviation is set to -1. The overlap between non-paired pointings was not taken into account for the search of duplicates, therefore the final catalogue will still contain two detections of the same source for $\sim$ 25\% of the sources. \\

\begin{figure}
\includegraphics[trim= 0.5cm 2cm 1cm 3cm, clip, width=9.2cm]{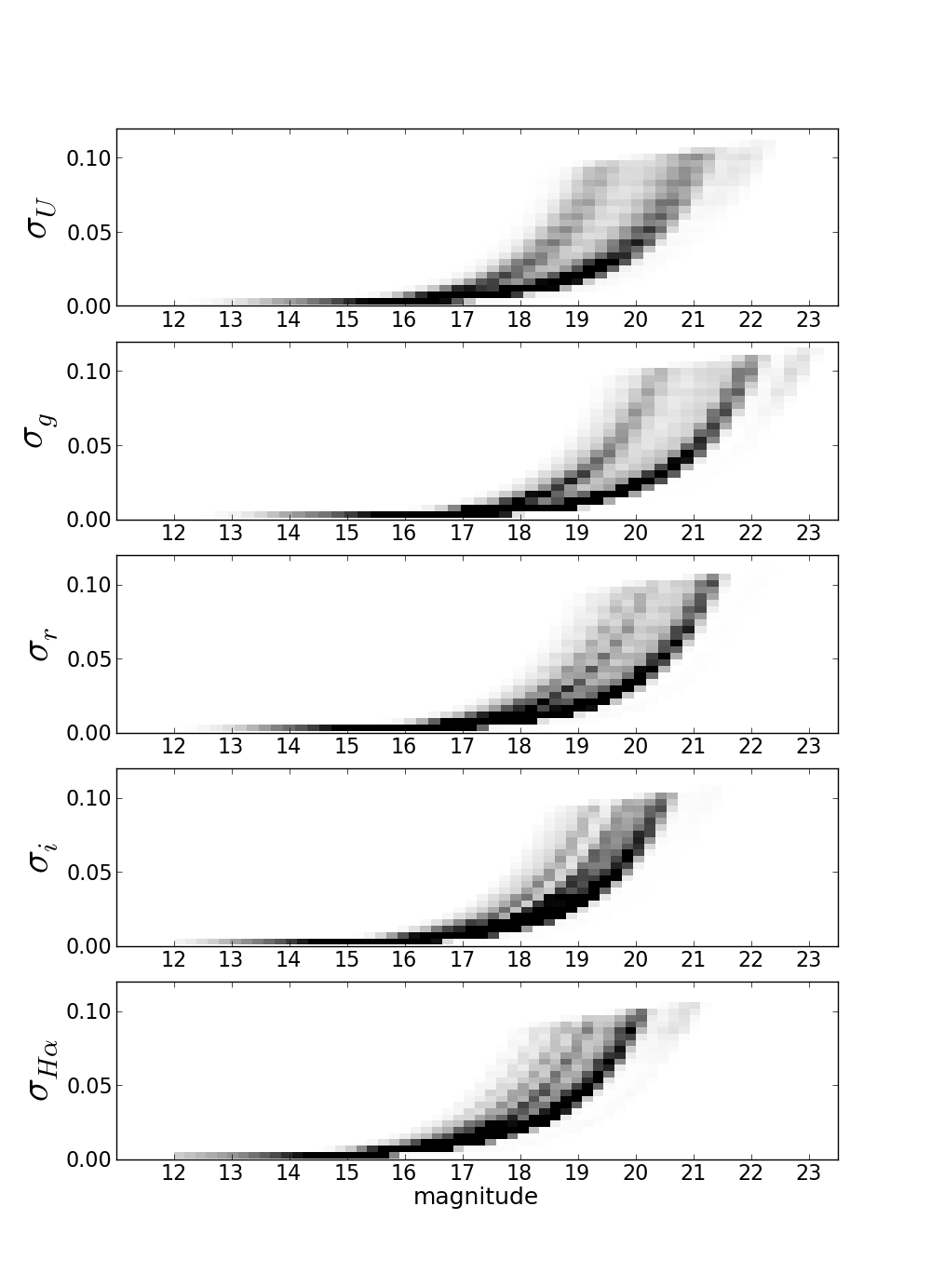}
\caption{Limiting magnitudes in all five bands. Sources with magnitudes smaller than 12$^{th}$ mag are detected but not shown because they are not considered reliable. The survey depth is $\sim$ 20$^{th}$ mag in all filters. We use a linear scale in order to show the densest regions of the plots. \label{limiting-mags}}
\end{figure}

 The limiting magnitudes in each filter can be seen in Figure \ref{limiting-mags}, where we plot only `stellar' objects in all five bands independently, for sources fainter than 12$^{th}$ mag in each filter. We also only plot significant detections, by setting the S/N threshold to values above $\sim$ 10, in order to avoid being misled by spurious detections. The figure shows that the depth of the ongoing survey is  $\sim$ 20$^{th}$ mag in the Vega system.  Sources brighter than $\sim$ 12$^{th}$ magnitude are saturated and their magnitudes should not be considered reliable. This figure also gives an indication of the photometric errors in the catalogue, as a function of magnitude. At $\sim$ 20$^{th}$ magnitude, the random photometric errors are $\sim$ 10\%, except in the case of the $U$-band where the errors can be larger. Errors of around 10\% are within the quality target of our catalogue. Note, the random photometric errors are obtained from the pipeline product and do not include systematic effects, e.g. related to the offset applied to calibrate KIS against KIC. The systematic errors (up to 5\%) may dominate at the bright end.
 
\begin{figure}
\includegraphics[trim= 1cm 0.5cm 0.5cm 1.5cm, clip, width=9.cm]{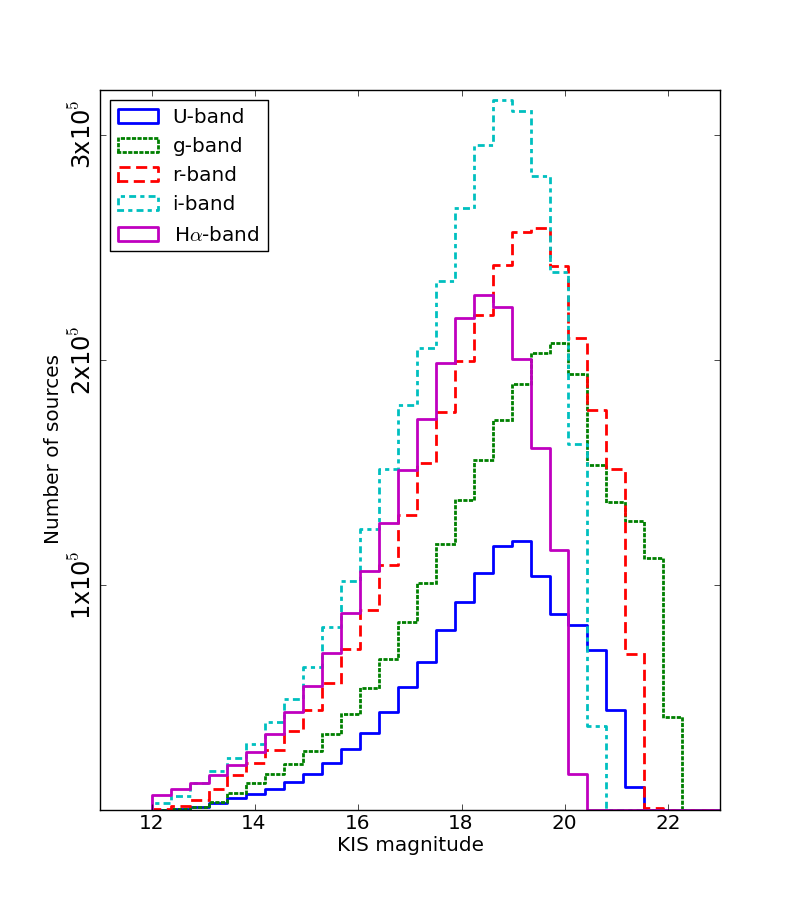}
\caption{Distribution of the number of sources as a function of KIS magnitudes. The objects taken into account are the ones classified as `stellar' in all five filters independently. \label{dist-mag}}
\end{figure}

We show the distribution of the number of detected sources as a function of magnitudes in Figure \ref{dist-mag}. Similarly to Figure \ref{limiting-mags}, we select `stellar' objects in all five bands independently and set the same S/N limit. As can be seen, the $g$ and $r$-bands go slightly deeper than the other filters and the number of detected objects in the $U$-band is much smaller than in the other bands.  \\

 In order to test the potential of the KIS catalogue to identify rare and unusual objects in the {\it Kepler} field, we cross-match KIS with known published sources within the {\it Kepler} FoV such as the pulsating subdwarfs and white dwarfs from \cite{ostensen10, ostensen11-1, ostensen11-2, hermes11}, the cataclysmic variables (CVs) from \cite{williamsetal10, woodetal11, fontaineetal11} and the active galactic nuclei (AGN) from \cite{mushotzkyetal11}. The objects which had matches in KIS are plotted in colour space (Figure \ref{ccd}). We also include a recently spectroscopically confirmed ultra-cool dwarf (R. Tata, E. L. Mart\'in \& E. Martioli, private communication) in the ($r $- H$\alpha$, $r - i$) colour-colour diagram of Figure \ref{ccd}, which did not have a $U$-band detection. We see that the different types of objects fall within their expected locations in colour-space. The CVs are indeed found in the ($U - g$, $g - r$) colour-colour diagram with ($U - g$) $<$ 0 and they also stand out as H$\alpha$ emitters in the ($r - H\alpha$, $r - i$) diagram. Also, the DA white dwarfs are known to be H$\alpha$ deficit objects due to strong broad Balmer line absorption, which can be seen in the bottom panel of Figure \ref{ccd}.
 
 We stress once again that all magnitudes are in the Vega system and the stellar model tracks can be found in the IPHAS and UVEX papers by \cite{drewetal05-1, grootetal09, verbeeketal12}. We also note that during all data analysis and manipulation steps, we always select KIS objects which have morphological classes in all five filters equal to -1, corresponding to stellar objects only. These amount to slightly under $\sim$ 1 million unique objects. \\

\section{Conclusion}

We obtained $Ugri$ and H$\alpha$ data for part of the {\it Kepler} field using the WFC on the INT. The data were processed at the Cambridge Astronomical Survey Unit (CASU) in the same fashion as for IPHAS \citep{drewetal05-1} and UVEX \citep{grootetal09}, with the exception of the $U$-band magnitudes, which in the KIS case were calculated using the mean ZP values from the standard stars observed each night. \\

 The KIS magnitudes were calibrated by shifting our zero-points to match the KIC $gri$ photometry. This way, we improved our photometric calibration on a pointing by pointing basis rather than across a full night. \\

Of the 742 pointings obtained throughout the 2011 season, 511 of them passed the quality control threshold set for this survey, covering $\sim$ 50 deg$^{2}$. Most sources in the KIS have two detections. In such cases, we derive a mean magnitude and error using the two values for each detected object. These average values are the ones provided in the KIS catalogue described in this document.

 The initial data release KIS catalogue contains $\sim$ 6 million objects, with $Ugri$ and H$\alpha$ magnitudes, down to $\sim$ 20$^{th}$ magnitude in the Vega system. Out of those $\sim$ 6 million sources, $\sim$ 1.2 million of them are classified as `stellar' in all five filters. The plan is to observe the remainder of the field in 2012, followed by a second data release containing all data. \\

\section*{Acknowledgments}

This paper makes use of data collected at the Isaac Newton Telescope, operated on the island of La Palma, by the Isaac Newton Group in the Spanish Observatorio del Roque de los Muchachos. The observations were processed by the Cambridge Astronomy Survey Unit (CASU) at the Institute of Astronomy, University of Cambridge.
We acknowledge the use of data taken from the {\it Kepler} Input Catalog (KIC), as well as data taken from the Sloan Digital Sky Survey. \\

D. Steeghs acknowledges a STFC Advanced Fellowship.

R. H. {\O}stensen acknowledges funding from the European Research Council under the European Community's Seventh Framework Programme (FP7/2007--2013)/ERC grant agreement N$^{\underline{\mathrm o}}$\,227224 ({\sc prosperity}), as well as from the Research Council of K.U.Leuven grant agreement GOA/2008/04.


\begin{thebibliography}{41}
\bibitem[{{Abazajian} {et~al.}(2009){Abazajian}, {Adelman-McCarthy}, {Ag{\"u}eros}, 
	{Allam}, {Allende Prieto}, {An}, {Anderson}, {Anderson}, {Annis}, {Bahcall}}]{abazajianetal09}
	{Abazajian}, K.~N., {Adelman-McCarthy}, J.~K., {Ag{\"u}eros}, M.~A., {et~al.} 2009, ApJS, 182, 543

\bibitem[{{Basri} {et~al.}(2011){Basri}, {Walkowicz}, {Batalha}, {Gilliland},
  {Jenkins}, {Borucki}, {Koch}, {Caldwell}, {Dupree}, {Latham}, {Marcy},
  {Meibom}, \& {Brown}}]{basrietal11}
{Basri}, G., {Walkowicz}, L.~M., {Batalha}, N., {et~al.} 2011, AJ, 141, 20

\bibitem[{{Benk{\H o}} {et~al.}(2010){Benk{\H o}}, {Kolenberg}, {Szab{\'o}},
  {Kurtz}, {Bryson}, {Bregman}, {Still}, {Smolec}, {Nuspl}, {Nemec},
  {Moskalik}, {Kopacki}, {Koll{\'a}th}, {Guggenberger}, {di Criscienzo},
  {Christensen-Dalsgaard}, {Kjeldsen}, {Borucki}, {Koch}, {Jenkins}, \& {van
  Cleve}}]{benkoetal10}
{Benk{\H o}}, J.~M., {Kolenberg}, K., {Szab{\'o}}, R., {et~al.} 2010, MNRAS,
  409, 1585

\bibitem[{{Bloemen} {et~al.}(2011){Bloemen}, {Marsh}, {{\O}stensen},
  {Charpinet}, {Fontaine}, {Degroote}, {Heber}, {Kawaler}, {Aerts}, {Green},
  {Telting}, {Brassard}, {G{\"a}nsicke}, {Handler}, {Kurtz}, {Silvotti}, {van
  Grootel}, {Lindberg}, {Pursimo}, {Wilson}, {Gilliland}, {Kjeldsen},
  {Christensen-Dalsgaard}, {Borucki}, {Koch}, {Jenkins}, \&
  {Klaus}}]{bloemenetal11}
{Bloemen}, S., {Marsh}, T.~R., {{\O}stensen}, R.~H., {et~al.} 2011, MNRAS, 410,
  1787

\bibitem[{{Borucki} {et~al.}(2010){Borucki}, {Koch}, {Basri}, {Batalha},
  {Brown}, {Caldwell}, {Caldwell}, {Christensen-Dalsgaard}, {Cochran},
  {DeVore}, {Dunham}, {Dupree}, {Gautier}, {Geary}, {Gilliland}, {Gould},
  {Howell}, {Jenkins}, {Kondo}, {Latham}, {Marcy}, {Meibom}, {Kjeldsen},
  {Lissauer}, {Monet}, {Morrison}, {Sasselov}, {Tarter}, {Boss}, {Brownlee},
  {Owen}, {Buzasi}, {Charbonneau}, {Doyle}, {Fortney}, {Ford}, {Holman},
  {Seager}, {Steffen}, {Welsh}, {Rowe}, {Anderson}, {Buchhave}, {Ciardi},
  {Walkowicz}, {Sherry}, {Horch}, {Isaacson}, {Everett}, {Fischer}, {Torres},
  {Johnson}, {Endl}, {MacQueen}, {Bryson}, {Dotson}, {Haas}, {Kolodziejczak},
  {Van Cleve}, {Chandrasekaran}, {Twicken}, {Quintana}, {Clarke}, {Allen},
  {Li}, {Wu}, {Tenenbaum}, {Verner}, {Bruhweiler}, {Barnes}, \&
  {Prsa}}]{boruckietal10}
{Borucki}, W.~J., {Koch}, D., {Basri}, G., {et~al.} 2010, Science, 327, 977

\bibitem[{{Brown} {et~al.}(2011){Brown}, {Latham}, {Everett}, \&
  {Esquerdo}}]{brownetal11}
{Brown}, T.~M., {Latham}, D.~W., {Everett}, M.~E., \& {Esquerdo}, G.~A. 2011,
  AJ, 142, 112

\bibitem[{{Chaplin} {et~al.}(2010){Chaplin}, {Appourchaux}, {Elsworth},
  {Garc{\'{\i}}a}, {Houdek}, {Karoff}, {Metcalfe}, {Molenda-{\.Z}akowicz},
  {Monteiro}, {Thompson}, {Brown}, {Christensen-Dalsgaard}, {Gilliland},
  {Kjeldsen}, {Borucki}, {Koch}, {Jenkins}, {Ballot}, {Basu}, {Bazot},
  {Bedding}, {Benomar}, {Bonanno}, {Brand{\~a}o}, {Bruntt}, {Campante},
  {Creevey}, {Di Mauro}, {Do{\u g}an}, {Dreizler}, {Eggenberger}, {Esch},
  {Fletcher}, {Frandsen}, {Gai}, {Gaulme}, {Handberg}, {Hekker}, {Howe},
  {Huber}, {Korzennik}, {Lebrun}, {Leccia}, {Martic}, {Mathur}, {Mosser},
  {New}, {Quirion}, {R{\'e}gulo}, {Roxburgh}, {Salabert}, {Schou}, {Sousa},
  {Stello}, {Verner}, {Arentoft}, {Barban}, {Belkacem}, {Benatti}, {Biazzo},
  {Boumier}, {Bradley}, {Broomhall}, {Buzasi}, {Claudi}, {Cunha}, {D'Antona},
  {Deheuvels}, {Derekas}, {Garc{\'{\i}}a Hern{\'a}ndez}, {Giampapa}, {Goupil},
  {Gruberbauer}, {Guzik}, {Hale}, {Ireland}, {Kiss}, {Kitiashvili},
  {Kolenberg}, {Korhonen}, {Kosovichev}, {Kupka}, {Lebreton}, {Leroy},
  {Ludwig}, {Mathis}, {Michel}, {Miglio}, {Montalb{\'a}n}, {Moya}, {Noels},
  {Noyes}, {Pall{\'e}}, {Piau}, {Preston}, {Roca Cort{\'e}s}, {Roth}, {Sato},
  {Schmitt}, {Serenelli}, {Silva Aguirre}, {Stevens}, {Su{\'a}rez}, {Suran},
  {Trampedach}, {Turck-Chi{\`e}ze}, {Uytterhoeven}, {Ventura}, \&
  {Wilson}}]{chaplinetal10}
{Chaplin}, W.~J., {Appourchaux}, T., {Elsworth}, Y., {et~al.} 2010, ApJ Lett.,
  713, L169

\bibitem[{{Corradi} {et~al.}(2008){Corradi}, {Rodr{\'{\i}}guez-Flores},
  {Mampaso}, {Greimel}, {Viironen}, {Drew}, {Lennon}, {Mikolajewska}, {Sabin},
  \& {Sokoloski}}]{corradietal08}
{Corradi}, R.~L.~M., {Rodr{\'{\i}}guez-Flores}, E.~R., {Mampaso}, A., {et~al.}
  2008, A\&A, 480, 409

\bibitem[{{Coughlin} {et~al.}(2011){Coughlin}, {L{\'o}pez-Morales}, {Harrison},
  {Ule}, \& {Hoffman}}]{coughlinetal11}
{Coughlin}, J.~L., {L{\'o}pez-Morales}, M., {Harrison}, T.~E., {Ule}, N., \&
  {Hoffman}, D.~I. 2011, AJ, 141, 78

\bibitem[{{Drew} {et~al.}(2008){Drew}, {Greimel}, {Irwin}, \&
  {Sale}}]{drewetal08}
{Drew}, J.~E., {Greimel}, R., {Irwin}, M.~J., \& {Sale}, S.~E. 2008, MNRAS,
  386, 1761

\bibitem[{{Drew} {et~al.}(2005){Drew}, {Greimel}, {Irwin}, {Aungwerojwit},
  {Barlow}, {Corradi}, {Drake}, {G{\"a}nsicke}, {Groot}, {Hales}, {Hopewell},
  {Irwin}, {Knigge}, {Leisy}, {Lennon}, {Mampaso}, {Masheder}, {Matsuura},
  {Morales-Rueda}, {Morris}, {Parker}, {Phillipps}, {Rodriguez-Gil}, {Roelofs},
  {Skillen}, {Sokoloski}, {Steeghs}, {Unruh}, {Viironen}, {Vink}, {Walton},
  {Witham}, {Wright}, {Zijlstra}, \& {Zurita}}]{drewetal05-1}
{Drew}, J.~E., {Greimel}, R., {Irwin}, M.~J., {et~al.} 2005, MNRAS, 362, 753

\bibitem[{{Everett} {et~al.}(2012){Everett}, {Howell}, \&
  {Kinemuchi}}]{everettetal12}
{Everett}, M.~E., {Howell}, S.~B., \& {Kinemuchi}, K. 2012, ArXiv e-prints

\bibitem[{{Fontaine} {et~al.}(2011){Fontaine}, {Brassard}, {Green},
  {Charpinet}, {Dufour}, {Hubeny}, {Steeghs}, {Aerts}, {Randall}, {Bergeron},
  {Guvenen}, {O'Malley}, {Van Grootel}, {{\O}stensen}, {Bloemen}, {Silvotti},
  {Howell}, {Baran}, {Kepler}, {Marsh}, {Montgomery}, {Oreiro}, {Provencal},
  {Telting}, {Winget}, {Zima}, {Christensen-Dalsgaard}, \&
  {Kjeldsen}}]{fontaineetal11}
{Fontaine}, G., {Brassard}, P., {Green}, E.~M., {et~al.} 2011, ApJ, 726, 92

\bibitem[{{Gonz{\'a}lez-Solares} {et~al.}(2008){Gonz{\'a}lez-Solares},
  {Walton}, {Greimel}, {Drew}, {Irwin}, {Sale}, {Andrews}, {Aungwerojwit},
  {Barlow}, {van den Besselaar}, {Corradi}, {G{\"a}nsicke}, {Groot}, {Hales},
  {Hopewell}, {Hu}, {Irwin}, {Knigge}, {Lagadec}, {Leisy}, {Lewis}, {Mampaso},
  {Matsuura}, {Moont}, {Morales-Rueda}, {Morris}, {Naylor}, {Parker}, {Prema},
  {Pyrzas}, {Rixon}, {Rodr{\'{\i}}guez-Gil}, {Roelofs}, {Sabin}, {Skillen},
  {Suso}, {Tata}, {Viironen}, {Vink}, {Witham}, {Wright}, {Zijlstra}, {Zurita},
  {Drake}, {Fabregat}, {Lennon}, {Lucas}, {Mart{\'{\i}}n}, {Phillipps},
  {Steeghs}, \& {Unruh}}]{gonzalez-solares08}
{Gonz{\'a}lez-Solares}, E.~A., {Walton}, N.~A., {Greimel}, R., {et~al.} 2008,
  MNRAS, 388, 89

\bibitem[{{Gonz{\'a}lez-Solares} {et~al.}(2011){Gonz{\'a}lez-Solares}, {Irwin},
  {McMahon}, {Hodgkin}, {Lewis}, {Walton}, {Jarvis}, {Marchetti}, {Oliver},
  {P{\'e}rez-Fournon}, {Siana}, {Surace}, \& {Vaccari}}]{gonzalez-solares11}
{Gonz{\'a}lez-Solares}, E.~A., {Irwin}, M., {McMahon}, R.~G., {et~al.} 2011,
  MNRAS, 416, 927

\bibitem[{{Groot} {et~al.}(2009){Groot}, {Verbeek}, {Greimel}, {Irwin},
  {Gonz{\'a}lez-Solares}, {G{\"a}nsicke}, {de Groot}, {Drew}, {Augusteijn},
  {Aungwerojwit}, {Barlow}, {Barros}, {van den Besselaar}, {Casares},
  {Corradi}, {Corral-Santana}, {Deacon}, {van Ham}, {Hu}, {Heber}, {Jonker},
  {King}, {Knigge}, {Mampaso}, {Marsh}, {Morales-Rueda}, {Napiwotzki},
  {Naylor}, {Nelemans}, {Oosting}, {Pyrzas}, {Pretorius},
  {Rodr{\'{\i}}guez-Gil}, {Roelofs}, {Sale}, {Schellart}, {Steeghs}, {Szyszka},
  {Unruh}, {Walton}, {Weston}, {Witham}, {Woudt}, \& {Zijlstra}}]{grootetal09}
{Groot}, P.~J., {Verbeek}, K., {Greimel}, R., {et~al.} 2009, MNRAS, 399, 323

\bibitem[{{Hermes} {et~al.}(2011){Hermes}, {Mullally}, {{\O}stensen},
  {Williams}, {Telting}, {Southworth}, {Bloemen}, {Howell}, {Everett}, \&
  {Winget}}]{hermes11}
{Hermes}, J.~J., {Mullally}, F., {{\O}stensen}, R.~H., {et~al.} 2011, ApJ
  Lett., 741, L16

\bibitem[{{Koester}, D.}(2010)]{koester10}
{Koester}, D. 2010, Mem. Soc. Astron. Ital., 81, 921

\bibitem[{{Llama} {et~al.}(2012){Llama}, {Jardine}, {Mackay}, \&
  {Fares}}]{llamaetal12}
{Llama}, J., {Jardine}, M., {Mackay}, D.~H., \& {Fares}, R. 2012, ArXiv
  e-prints

\bibitem[{{Loveday}(2002)}]{loveday02}
{Loveday}, J. 2002, Contemporary Physics, 43, 437

\bibitem[{{Meibom} {et~al.}(2011){Meibom}, {Barnes}, {Latham}, {Batalha},
  {Borucki}, {Koch}, {Basri}, {Walkowicz}, {Janes}, {Jenkins}, {Van Cleve},
  {Haas}, {Bryson}, {Dupree}, {Furesz}, {Szentgyorgyi}, {Buchhave}, {Clarke},
  {Twicken}, \& {Quintana}}]{meibometal11}
{Meibom}, S., {Barnes}, S.~A., {Latham}, D.~W., {et~al.} 2011, ApJ Lett., 733,
  L9

\bibitem[{{Morgan} {et~al.}(1953){Morgan}, {Harris}, \&
  {Johnson}}]{morgan+johnson53}
{Morgan}, W.~W., {Harris}, D.~L., \& {Johnson}, H.~L. 1953, ApJ, 118, 92

\bibitem[{{Mushotzky} {et~al.}(2011){Mushotzky}, {Edelson}, {Baumgartner}, \&
  {Gandhi}}]{mushotzkyetal11}
{Mushotzky}, R.~F., {Edelson}, R., {Baumgartner}, W., \& {Gandhi}, P. 2011, ApJ
  Lett., 743, L12

\bibitem[{{Nemec} {et~al.}(2011){Nemec}, {Smolec}, {Benk{\H o}}, {Moskalik},
  {Kolenberg}, {Szab{\'o}}, {Kurtz}, {Bryson}, {Guggenberger}, {Chadid},
  {Jeon}, {Kunder}, {Layden}, {Kinemuchi}, {Kiss}, {Poretti},
  {Christensen-Dalsgaard}, {Kjeldsen}, {Caldwell}, {Ripepi}, {Derekas},
  {Nuspl}, {Mullally}, {Thompson}, \& {Borucki}}]{nemecetal11}
{Nemec}, J.~M., {Smolec}, R., {Benk{\H o}}, J.~M., {et~al.} 2011, MNRAS, 417,
  1022

\bibitem[{{Oke} \& {Gunn}(1983)}]{oke+gunn83}
{Oke}, J.~B., \& {Gunn}, J.~E. 1983, ApJ, 266, 713

\bibitem[{{{\O}stensen} {et~al.}(2011{\natexlab{a}}){{\O}stensen}, {Bloemen},
  {Vu{\v c}kovi{\'c}}, {Aerts}, {Oreiro}, {Kinemuchi}, {Still}, \&
  {Koester}}]{ostensen11-2}
{{\O}stensen}, R.~H., {Bloemen}, S., {Vu{\v c}kovi{\'c}}, M., {et~al.}
  2011{\natexlab{a}}, ApJ Lett., 736, L39

\bibitem[{{{\O}stensen} {et~al.}(2010){{\O}stensen}, {Silvotti}, {Charpinet},
  {Oreiro}, {Handler}, {Green}, {Bloemen}, {Heber}, {G{\"a}nsicke}, {Marsh},
  {Kurtz}, {Telting}, {Reed}, {Kawaler}, {Aerts}, {Rodr{\'{\i}}guez-L{\'o}pez},
  {Vu{\v c}kovi{\'c}}, {Ottosen}, {Liimets}, {Quint}, {van Grootel}, {Randall},
  {Gilliland}, {Kjeldsen}, {Christensen-Dalsgaard}, {Borucki}, {Koch}, \&
  {Quintana}}]{ostensen10}
{{\O}stensen}, R.~H., {Silvotti}, R., {Charpinet}, S., {et~al.} 2010, MNRAS,
  409, 1470

\bibitem[{{{\O}stensen} {et~al.}(2011{\natexlab{b}}){{\O}stensen}, {Silvotti},
  {Charpinet}, {Oreiro}, {Bloemen}, {Baran}, {Reed}, {Kawaler}, {Telting},
  {Green}, {O'Toole}, {Aerts}, {G{\"a}nsicke}, {Marsh}, {Breedt}, {Heber},
  {Koester}, {Quint}, {Kurtz}, {Rodr{\'{\i}}guez-L{\'o}pez}, {Vu{\v
  c}kovi{\'c}}, {Ottosen}, {Frimann}, {Somero}, {Wilson}, {Thygesen},
  {Lindberg}, {Kjeldsen}, {Christensen-Dalsgaard}, {Allen}, {McCauliff}, \&
  {Middour}}]{ostensen11-1}
---. 2011{\natexlab{b}}, MNRAS, 414, 2860

\bibitem[{{Pickles}, A.~J.}(1998){Pickles}]{pickles98}
{Pickles}, A.~J. 1998, PASP, 110, 863

\bibitem[{{Pr{\v s}a} {et~al.}(2011){Pr{\v s}a}, {Batalha}, {Slawson}, {Doyle},
  {Welsh}, {Orosz}, {Seager}, {Rucker}, {Mjaseth}, {Engle}, {Conroy},
  {Jenkins}, {Caldwell}, {Koch}, \& {Borucki}}]{prsaetal11}
{Pr{\v s}a}, A., {Batalha}, N., {Slawson}, R.~W., {et~al.} 2011, AJ, 141, 83

\bibitem[{{Scaringi} {et~al.}(2012){Scaringi}, {Kording}, {Uttley}, {Knigge},
  {Groot}, \& {Still}}]{scaringietal12}
{Scaringi}, S., {Kording}, E., {Uttley}, P., {et~al.} 2012, ArXiv e-prints

\bibitem[{{Skrutskie} {et~al.}(2006){Skrutskie}, {Cutri}, {Stiening},
  {Weinberg}, {Schneider}, {Carpenter}, {Beichman}, {Capps}, {Chester},
  {Elias}, {Huchra}, {Liebert}, {Lonsdale}, {Monet}, {Price}, {Seitzer},
  {Jarrett}, {Kirkpatrick}, {Gizis}, {Howard}, {Evans}, {Fowler}, {Fullmer},
  {Hurt}, {Light}, {Kopan}, {Marsh}, {McCallon}, {Tam}, {Van Dyk}, \&
  {Wheelock}}]{skrutskieetal06-1}
{Skrutskie}, M.~F., {Cutri}, R.~M., {Stiening}, R., {et~al.} 2006, AJ, 131,
  1163

\bibitem[{{Still} {et~al.}(2010){Still}, {Howell}, {Wood}, {Cannizzo}, \&
  {Smale}}]{stilletal10}
{Still}, M., {Howell}, S.~B., {Wood}, M.~A., {Cannizzo}, J.~K., \& {Smale},
  A.~P. 2010, ApJ Lett., 717, L113

\bibitem[{{Stoughton} {et~al.}(2002){Stoughton}, {Lupton}, {Bernardi},
  {Blanton}, {Burles}, {Castander}, {Connolly}, {Eisenstein}, {Frieman},
  {Hennessy}, {Hindsley}, {Ivezi{\' c}}, {Kent}, {Kunszt}, {Lee}, {Meiksin},
  {Munn}, {Newberg}, {Nichol}, {Nicinski}, {Pier}, {Richards}, {Richmond},
  {Schlegel}, {Smith}, {Strauss}, {SubbaRao}, {Szalay}, {Thakar}, {Tucker},
  {Vanden Berk}, {Yanny}, \& {...}}]{stoughtonetal02-1}
{Stoughton}, C., {Lupton}, R.~H., {Bernardi}, M., {et~al.} 2002, AJ, 123, 485

\bibitem[{{Valdivielso} {et~al.}(2009){Valdivielso}, {Mart{\'{\i}}n}, {Bouy},
  {Solano}, {Drew}, {Greimel}, {Guti{\'e}rrez}, {Unruh}, \&
  {Vink}}]{valdivielsoetal09}
{Valdivielso}, L., {Mart{\'{\i}}n}, E.~L., {Bouy}, H., {et~al.} 2009, A\&A,
  497, 973

\bibitem[{{van Kerkwijk} {et~al.}(2010){van Kerkwijk}, {Rappaport}, {Breton},
  {Justham}, {Podsiadlowski}, \& {Han}}]{vankerkwijketal10}
{van Kerkwijk}, M.~H., {Rappaport}, S.~A., {Breton}, R.~P., {et~al.} 2010, ApJ,
  715, 51

\bibitem[{{Verbeek} {et~al.}(2012){Verbeek}, {de Groot}, {Groot}, {Scaringi},
  {Drew}, {Greimel}, {Irwin}, {Gonz{\'a}lez-Solares}, {G{\"a}nsicke},
  {Casares}, {Corral-Santana}, {Deacon}, \& {Steeghs}}]{verbeeketal12}
{Verbeek}, K., {de Groot}, E., {Groot}, P.~J., {et~al.} 2012, MNRAS, 420, 1115

\bibitem[{{Viironen} {et~al.}(2009){Viironen}, {Greimel}, {Corradi}, {Mampaso},
  {Rodr{\'{\i}}guez}, {Sabin}, {Delgado-Inglada}, {Drew}, {Giammanco},
  {Gonz{\'a}lez-Solares}, {Irwin}, {Miszalski}, {Parker},
  {Rodr{\'{\i}}guez-Flores}, \& {Zijlstra}}]{viironenetal09}
{Viironen}, K., {Greimel}, R., {Corradi}, R.~L.~M., {et~al.} 2009, A\&A, 504,
  291

\bibitem[{{Williams} {et~al.}(2010){Williams}, {de Martino}, {Silvotti},
  {Bruni}, {Dufour}, {Riecken}, {Kronberg}, {Mukadam}, \&
  {Handler}}]{williamsetal10}
{Williams}, K.~A., {de Martino}, D., {Silvotti}, R., {et~al.} 2010, ApJ, 139,
  2587

\bibitem[{{Witham} {et~al.}(2006){Witham}, {Knigge}, {G{\"a}nsicke},
  {Aungwerojwit}, {Corradi}, {Drew}, {Greimel}, {Groot}, {Morales-Rueda},
  {Rodriguez-Flores}, {Rodriguez-Gil}, \& {Steeghs}}]{withametal06-1}
{Witham}, A.~R., {Knigge}, C., {G{\"a}nsicke}, B.~T., {et~al.} 2006, MNRAS,
  369, 581

\bibitem[{{Witham} {et~al.}(2007){Witham}, {Knigge}, {Aungwerojwit}, {Drew},
  {G{\"a}nsicke}, {Greimel}, {Groot}, {Roelofs}, {Steeghs}, \&
  {Woudt}}]{withametal07}
{Witham}, A.~R., {Knigge}, C., {Aungwerojwit}, A., {et~al.} 2007, MNRAS, 382,
  1158

\bibitem[{{Wood} {et~al.}(2011){Wood}, {Still}, {Howell}, {Cannizzo}, \&
  {Smale}}]{woodetal11}
{Wood}, M.~A., {Still}, M.~D., {Howell}, S.~B., {Cannizzo}, J.~K., \& {Smale},
  A.~P. 2011, ApJ, 741, 105

\bibitem[{{Wright} {et~al.}(2009){Wright}, {Barlow}, {Greimel}, {Drew},
  {Matsuura}, {Unruh}, \& {Zijlstra}}]{wrightetal09}
{Wright}, N.~J., {Barlow}, M.~J., {Greimel}, R., {et~al.} 2009, MNRAS, 400,
  1413

\bibitem[{{Wright} {et~al.}(2008){Wright}, {Greimel}, {Barlow}, {Drew},
  {Cioni}, {Zijlstra}, {Corradi}, {Gonz{\'a}lez-Solares}, {Groot}, {Irwin},
  {Irwin}, {Mampaso}, {Morris}, {Steeghs}, {Unruh}, \& {Walton}}]{wrightetal08}
{Wright}, N.~J., {Greimel}, R., {Barlow}, M.~J., {et~al.} 2008, MNRAS, 390, 929

\end{thebibliography}

\clearpage
\begin{table}
\caption{Description of columns in KIS catalogue \label{col-desc}}
\small
\setstretch{1.8}
\begin{tabular}{l l}
\hline \hline
Column name & Description\\
\hline
KIS\_ID \dotfill & KIS ID containing the coordinate of the source, in \\
 & sexagesimal notation. \\
RA \dotfill & Right Ascension (J2000), in degrees. \\
Dec \dotfill & Declination (J2000), in degrees. \\
mean\_$U$, mean\_$g$, mean\_$r$, mean\_$i$, mean\_H$\alpha$ \dotfill & Magnitudes of sources, given in the Vega system. \\
 & In the case of two detections, the mean value is given. \\
$U$\_err, $g$\_err, $r$\_err, $i$\_err, H$\alpha$\_err \dotfill & Magnitude errors. In the case of two detections, the \\
 & mean error is given. \\
rms\_$U$\tablenotemark{$\star$}, rms\_$g$\tablenotemark{$\star$}, rms\_$r$\tablenotemark{$\star$}, rms\_$i$\tablenotemark{$\star$}, rms\_H$\alpha$\tablenotemark{$\star$} \dotfill & Root-mean-square (rms) deviation of magnitudes of sources \\
 & with two detections. In the case of a single detection, the \\
  & rms deviation value is set to -1. \\
x\_$U$\tablenotemark{$\star$}, x\_$g$\tablenotemark{$\star$}, x\_$r$\tablenotemark{$\star$}, x\_$i$\tablenotemark{$\star$}, x\_H$\alpha$\tablenotemark{$\star$} \dotfill & X pixel coordinate of source. \\
y\_$U$\tablenotemark{$\star$}, y\_$g$\tablenotemark{$\star$}, y\_$r$\tablenotemark{$\star$}, y\_$i$\tablenotemark{$\star$}, y\_H$\alpha$\tablenotemark{$\star$} \dotfill & Y pixel coordinate of source. \\
class\_$U$, class\_$g$, class\_$r$, class\_$i$, class\_H$\alpha$ \dotfill & Morphological class of source (see Table \ref{morph}). \\
CCD\tablenotemark{$\star$} & WFC's CCD in which the source was detected. \\
seeing\_$U$\tablenotemark{$\star$}, seeing\_$g$\tablenotemark{$\star$}, seeing\_$r$\tablenotemark{$\star$}, seeing\_$i$\tablenotemark{$\star$}, seeing\_H$\alpha$\tablenotemark{$\star$} \dotfill & Average seeing of the INT pointing, given in arcsec. \\
ellipticity\_$U$\tablenotemark{$\star$}, ellipticity\_$g$\tablenotemark{$\star$}, ellipticity\_$r$\tablenotemark{$\star$}, ellipticity\_$i$\tablenotemark{$\star$}, ellipticity\_H$\alpha$\tablenotemark{$\star$} \dotfill & Average ellipticity of the night. \\
MJD\_$U$\tablenotemark{$\star$}, MJD\_$g$\tablenotemark{$\star$}, MJD\_$r$\tablenotemark{$\star$}, MJD\_$i$\tablenotemark{$\star$}, MJD\_H$\alpha$\tablenotemark{$\star$} \dotfill & Modified julian date of observation. \\
delta\_$U$\tablenotemark{$\star$}, delta\_$g$\tablenotemark{$\star$}, delta\_$r$\tablenotemark{$\star$}, delta\_$i$\tablenotemark{$\star$}, delta\_H$\alpha$\tablenotemark{$\star$} \dotfill & Difference between KIS and KIC magnitudes applied to \\
 & calibrate KIS data. \\
KIC\_ID \dotfill & KIC ID of source. If it does not have a KIC match within \\
 & 1 arcsec, the value is set to 0. \\
\hline
\end{tabular}
$^\star$ These values are only available in the extended catalogue.
\end{table}

\begin{sidewaystable}
\setstretch{1.8}
\caption{Example of light version of the KIS catalogue \label{cat}}
\resizebox {1.\textwidth}{!}{%
\begin{tabular}{c c c c c c c c c c c c c c c c c c c}
\hline \hline
KIS ID & RA & Dec & $U$ & $\sigma_u$ & cls $U$ & $g$ & $\sigma_g$ & cls $g$ & $r$ & $\sigma_r$ & cls $r$ & $i$ & $\sigma_i$ & cls $i$ & H$\alpha$ & $\sigma_{H\alpha}$ & cls H$\alpha$ & KIC ID\\
\hline
KISJ183904.45+472458.6 & 279.768542 & 47.416278 & 20.892 & 0.150 & -1 & 18.811 & 0.014 & -1 & 17.754 & 0.000 & -1 & 17.110 & 0.000 & -1 & 0.000 & 0.000 &  0 & 10316732 \\
KISJ183904.53+471612.6 & 279.768875 & 47.270167 & 0.000 & 0.000 &  0 & 21.395 & 0.120 & -1 & 0.000 & 0.000 &  0 & 0.000 & 0.000 &  0 & 0.000 & 0.000 &  0 & 0 \\
KISJ183904.68+471532.8 & 279.769500 & 47.259111 & 0.000 & 0.000 &  0 & 13.499 & 0.001 & -1 & 12.888 & 0.001 & -9 & 12.489 & 0.001 & -1 & 0.000 & 0.000 &  0 & 0 \\
KISJ183904.68+471625.6 & 279.769500 & 47.273778 & 0.000 & 0.000 &  0 & 21.955 & 0.193 & -1 & 20.490 & 0.073 &  1 & 18.921 & 0.033 &  1 & 19.699 & 0.091 &  1 & 0 \\
KISJ183904.73+473001.1 & 279.769708 & 47.500306 & 0.000 & 0.000 &  0 & 0.000 & 0.000 &  0 & 21.326 & 0.148 & -1 & 19.421 & 0.050 & -1 & 20.253 & 0.145 & -1 & 0 \\
KISJ183904.76+471636.7 & 279.769833 & 47.276861 & 0.000 & 0.000 &  0 & 0.000 & 0.000 &  0 & 21.789 & 0.219 & -1 & 20.332 & 0.110 &  1 & 0.000 & 0.000 &  0 & 0 \\
KISJ183904.79+472829.3 & 279.769958 & 47.474806 & 18.758 & 0.025 & -1 & 18.042 & 0.000 & -1 & 17.274 & 0.006 & -1 & 16.821 & 0.007 & -1 & 16.964 & 0.011 & -1 & 10316736 \\
KISJ183904.85+472027.0 & 279.770208 & 47.340833 & 18.010 & 0.014 & -1 & 17.571 & 0.006 & -1 & 16.931 & 0.005 & -1 & 16.486 & 0.006 & -1 & 16.623 & 0.009 & -1 & 10250924 \\
KISJ183904.88+473120.4 & 279.770333 & 47.522333 & 16.383 & 0.005 & -1 & 0.000 & 0.000 &  0 & 15.481 & 0.002 & -1 & 0.000 & 0.000 &  0 & 0.000 & 0.000 &  0 & 10382031\\
KISJ183904.93+472511.8 & 279.770542 & 47.419944 & 19.668 & 0.053 & -1 & 18.530 & 0.011 & -1 & 17.582 & 0.007 & -1 & 16.981 & 0.000 & -1 & 17.181 & 0.013 & -1 & 10316737\\
KISJ183904.97+470905.8 & 279.770708 & 47.151611 & 0.000 & 0.000 &  0 & 21.745 & 0.160 & -7 & 0.000 & 0.000 &  0 & 0.000 & 0.000 &  0 & 0.000 & 0.000 &  0 & 0 \\
KISJ183904.98+471902.8 & 279.770750 & 47.317444 & 12.859 & 0.001 & -9 & 11.694 & 0.001 & -9 & 10.531 & 0.001 & -9 & 10.252 & 0.001 & -9 & 10.519 & 0.001 & -9 & 0 \\
KISJ183905.04+472525.8 & 279.771000 & 47.423833 & 0.000 & 0.000 &  0 & 0.000 & 0.000 &  0 & 21.540 & 0.178 & -1 & 19.505 & 0.054 &  1 & 0.000 & 0.000 &  0 & 0 \\
KISJ183915.68+472135.5 & 279.815333 & 47.359861 & 15.487 & 0.003 & -1 & 14.960 & 0.001 & -1 & 14.315 & 0.001 & -1 & 13.880 & 0.001 & -1 & 14.024 & 0.002 & -1 & 10250986 \\
KISJ183915.74+472239.7 & 279.815583 & 47.377694 & 0.000 & 0.000 &  0 & 0.000 & 0.000 &  0 & 21.466 & 0.168 & -1 & 21.016 & 0.197 &  1 & 0.000 & 0.000 &  0 & 0 \\
KISJ183915.77+472054.0 & 279.815708 & 47.348333 & 19.701 & 0.055 & -1 & 18.899 & 0.015 & -1 & 18.130 & 0.011 & -1 & 17.598 & 0.012 & -1 & 17.759 & 0.020 & -1 & 10250987 \\
KISJ183916.12+472412.8 & 279.817167 & 47.403556 & 19.686 & 0.054 & -1 & 18.102 & 0.000 & -1 & 17.000 & 0.005 & -1 & 16.390 & 0.005 & -1 & 16.613 & 0.009 & -1 & 10316799 \\
KISJ183916.16+472140.4 & 279.817333 & 47.361222 & 16.751 & 0.006 & -1 & 16.443 & 0.003 & -1 & 15.871 & 0.003 & -1 & 15.459 & 0.003 & -1 & 15.597 & 0.005 & -1 & 10250990 \\
KISJ1839163+472143.4 & 279.817625 & 47.362056 & 19.078 & 0.032 &  1 & 18.633 & 0.012 &  1 & 18.026 & 0.010 & -1 & 17.560 & 0.011 & -1 & 17.720 & 0.019 & -1 & 0 \\
KISJ183916.33+472123.7 & 279.818042 & 47.356583 & 19.360 & 0.041 & -1 & 18.687 & 0.013 & -1 & 17.957 & 0.010 & -1 & 17.428 & 0.010 & -1 & 17.601 & 0.017 & -1 & 10250993 \\
KISJ183916.35+472130.3 & 279.818125 & 47.358417 & 20.833 & 0.144 & -1 & 18.616 & 0.012 & -1 & 17.088 & 0.005 & -1 & 15.156 & 0.003 & -1 & 16.070 & 0.007 & -1 & 0 \\
KISJ183916.40+471515.6 & 279.818333 & 47.254333 & 15.715 & 0.003 & -1 & 14.878 & 0.001 & -1 & 14.131 & 0.001 & -1 & 13.675 & 0.001 & -1 & 13.812 & 0.002 & -1 & 10184560 \\
KISJ183916.47+471621.6 & 279.818625 & 47.272667 & 18.240 & 0.017 & -1 & 16.727 & 0.004 & -1 & 15.673 & 0.002 & -1 & 15.065 & 0.002 & -1 & 15.274 & 0.004 & -1 & 10184561 \\
KISJ183916.79+471239.7 & 279.819958 & 47.211028 & 20.413 & 0.100 & -1 & 20.341 & 0.049 & -2 & 19.823 & 0.041 & -1 & 19.318 & 0.046 &  1 & 19.494 & 0.077 & -1 & 10184563 \\
KISJ183916.83+472018.8 & 279.820125 & 47.338556 & 19.927 & 0.066 & -1 & 19.143 & 0.018 & -1 & 18.330 & 0.013 & -1 & 17.758 & 0.013 & -1 & 18.000 & 0.023 & -1 & 10250996 \\
\hline
\end{tabular}}
\end{sidewaystable}

\end{document}